\documentclass[a4paper,fleqn]{cas-dc}
\usepackage[authoryear]{natbib}
\usepackage[utf8]{inputenc}
\usepackage{amsmath}
\usepackage{mathtools}
\usepackage{placeins}
\usepackage{subfigure}
\usepackage{caption}
\begin{document}
\let\WriteBookmarks\relax
\def\floatpagepagefraction{1}
\def\textpagefraction{.001}
\shortauthors{Anis Moradikouchi et~al.}
\title [mode = title]{Terahertz Frequency-Domain Sensing Combined with Quantitative Multivariate Analysis for  Pharmaceutical Tablet Inspection}                      

\author[1,2]{Anis Moradikouchi}[
                        orcid=0000-0003-3811-1024]
\cormark[1]

\credit{Conceptualization of this study, Methodology, Software}

\address[1]{Department of Microtechnology and Nanoscience, Chalmers University of Technology, SE-412 96 Gothenburg, Sweden}
\author[2]{Anders Sparén}
\author [2]{Olof Svensson}
\author[3]{Staffan Folestad}
\author[1]{Jan Stake}
\author[1]{Helena Rodilla}

\credit{Data curation, Writing - Original draft preparation}

\address[2]{Oral Product Development, Pharmaceutical Technology \& Development, Operations, AstraZeneca, Gothenburg, Sweden}

\address[3]{Innovation Strategies \& External Liaison, Pharmaceutical Technology \& Development, Operations, AstraZeneca, Gothenburg, Sweden}

\cortext[cor1]{anismo@chalmers.se}

\pagenumbering{arabic}

\shorttitle{}



\maketitle 

\begin{abstract}
Near infrared (NIR) and Raman spectroscopy combined with multivariate analysis are established techniques for the identification and quantification of chemical properties of pharmaceutical tablets like the concentration of active pharmaceutical ingredients (API). However, these techniques suffer from a high sensitivity to particle size variations and are not ideal for the characterization of physical properties of tablets such as tablet density. In this work, we have explored the feasibility of terahertz frequency-domain spectroscopy, with the advantage of low scattering effects, combined with multivariate analysis to quantify API concentration and tablet density.  We studied 33 tablets, consisting of Ibuprofen, Mannitol, and a lubricant with API concentration and filler particle size as the design factors. The terahertz signal was measured in transmission mode across the frequency range 750 GHz to 1.5 THz using a vector
network analyzer, frequency extenders, horn antennas, and four off-axis parabolic mirrors. The attenuation spectral data were pre-processed and orthogonal partial least square (OPLS) regression was applied to the spectral data to obtain quantitative prediction models for API concentration and tablet density. The performance of the models was assessed using test sets. While a fair model was obtained for API concentration, a high-quality model was demonstrated for tablet density. The coefficient of determination ($R^2$) for the calibration set was 0.97 for tablet density and 0.98 for API concentration, while the relative prediction errors for the test set were 0.7\% and 6\% for tablet density and API concentration models, respectively. In conclusion, terahertz spectroscopy demonstrated to be a complementary technique to Raman and  NIR spectroscopy, which enables the characterization of physical properties of tablets like tablet density, and the characterization of API concentration with the advantage of low scattering effects.
\end{abstract}
\section{Introduction}
The pharmaceutical industry needs new analytical tools for real-time  monitoring and control of their products in the manufacturing area specifically suited for continuous manufacturing. Traditionally, in batch manufacturing, random samples of the finished products are transported to a laboratory to analyse and verify their quality. This process is slow and costly, with a potential risk that the final product should not meet its specifications, leading to batch failure. In this context, there has been a shift from the batch manufacturing process to continuous manufacturing~\citep{Plumb2005ContinuousSet, Ierapetritou2016PerspectivesProcesses}, where there is an increasing need for process analytical technologies that can be integrated into the manufacturing line and allow advanced process monitoring and control of the process and product. 

Over the years, scientific and technological progress has led to several new techniques  for the analysis of chemical and physical properties of drug products. Vibrational spectroscopy includes a series of key technologies that allow non-destructive, non-invasive, and rapid analysis of pharmaceutical materials~\citep{Ewing2018RecentFormulations}. NIR and Raman spectroscopy are established reliable techniques in process analytical technology for material analysis based on probing molecular structure and interactions. The use of these techniques combined with multivariate analysis such as partial-least-squares (PLS) and principal component analysis enables  qualitative and quantitative analysis of pharmaceutical materials~\citep{Laske2017AMolecules}.
NIR spectroscopy in diffuse reflection mode has been extensively used 
for material quantification~\citep{Berntsson2002QuantitativeSpectroscopy} and probing blend uniformity~\citep{Lyon2002Near-infraredHomogeneity}.
However, this technique suffers from high scattering effects and low sample penetration depth~\citep{Berntsson1999EffectiveSpectrometry}, which restricts the
extracted information to the surface properties. Moreover, its application for the analysis of low dosage material is limited~\citep{Sparen2015MatrixSpectroscopy}. 
Raman spectroscopy has been used for synthesis  monitoring~\citep{Svensson2000TheChemometrics}, blend monitoring~\citep{Vergote2004In-lineSpectroscopy},  determination of polymorphs~\citep{Aina2010TransmissionFormulations}, and quantification of drugs~\citep{Jedvert1998QuantificationSpectroscopy}. Raman spectroscopy offers more chemical specificity than NIR due to probing the fundamental vibrational modes. However, this technique is sensitive to  particle size variations due to high scattering effects~\citep{Townshend2012EffectScattering}, and if the laser power is too high, long exposure could burn the samples. Low-frequency Raman spectroscopy provides access to the lower frequencies from 300 GHz to 6 THz, but performs poorly when applied to compounds with strong fluorescence~\citep{Salim2020Low-frequencyDigestion}.

Terahertz spectroscopy~\citep{Siegel2004TerahertzMedicine} is a complementary technique to the spectral information from NIR and Raman spectroscopy for the characterization and identification of materials~\citep{Bawuah2021AdvancesReviewb}. Spectroscopy in the terahertz region, 300 GHz to 10 THz, identifies the low frequency vibrations associated with intermolecular interactions and the morphology of formulated drugs~\citep{Claybourn2007TerahertzApplicationsb}. Moreover, Terahertz spectroscopy enables the characterization of the physical properties of materials like tablet density and porosity~\citep{Bawuah2020Terahertz-BasedTutorial, Moradikouchi2022TerahertzTablets}. Taday~\citep{Taday2004ApplicationsSciences} demonstrated the use of terahertz pulsed spectroscopy for the quantification of paracetamol and aspirin tablets using a PLS calibration model. \textcolor{black}{In a study by Zeitler et. al. \citep{Zeitler2007DrugSpectroscopy}, combined with principle component analysis, the sensitivity of terahertz pulsed spectroscopy to anhydrous and hydrated pharmaceutical materials was investigated.}   Hisazumi et al.~\citep{Hisazumi2012UsingTablets} demonstrated the applicability of terahertz reflectance spectroscopy with partial least squares regression for the quantification of API in tablets. Yang et al.~\citep{Yang2021QualitativeMethod} used terahertz time-domain spectroscopy (THz-TDS) combined with support vector regression chemometric method to analyse the characteristic spectrum of caffeine in medicine. Peng et al.~\citep{Peng2018QualitativeSpectroscopy} proposed THz-TDS combined with support vector regression chemometric method to characterize substances in brain glioma. 

In this paper, we propose terahertz frequency domain spectroscopy (THz-FDS), based on \textcolor{black}{ electronic heterodyne techniques \citep{Hubers2008TerahertzReceivers}}, in combination with multivariate analysis for the analysis of API concentration and tablet density of pharmaceutical tablets as a complementary technique to NIR and Raman spectroscopy~\citep{Sparen2015MatrixSpectroscopy}. The high frequency resolution and stability of electronic heterodyne techniques minimizes the propagation of systematic errors when performing reference measurements~\citep{Hubers2011TerahertzConsiderations}. A quantitative analysis of the tablets was performed including the following steps. First, the complex transmission coefficients in phase and amplitude ($S_{21}$) ~\citep{Pozar2012MicrowaveEdition} of the tablets were measured and the attenuation spectra were obtained. Then, the spectral data were pre-processed to remove noise and baseline variations. Orthogonal partial least squares (OPLS) regression was applied to the spectral data to establish a prediction model for the tablets. Finally, an independent test set was used to assess the performance of the prediction models for API concentration and tablet density.

\section{Methodology}

\subsection{ Tablet preparation}
We studied tablets consisting of Ibuprofen as an API, Mannitol as a filler, and Magnesium stearate as a lubricant.  A fraction of a full factorial design from reference~\citep{Sparen2015MatrixSpectroscopy} was used for our experiments. \textcolor{black}{The API, filler, and lubricant were blended in a Turbula 2TF blender (Glen Mills
Inc., Switzerland). Tablets were manufactured through direct compaction with a single punch press
Korsch EK-0 (Korsch AG, Germany) equipped with flat
round 10 mm punches.} The design factors were Mannitol particle size, varied at two levels ($d_{50}$ $\sim$ 91, 450 $\mu$m), and API concentration varied at five levels ($\sim$ 16, 18, 20, 22, and 24 w/w \%). \textcolor{black}{Tablets with the applied compaction force of 12 kN were used for the experiments.} The Magnesium stearate concentration was constant at $\sim$ 1 w/w\% for all tablets. As shown in table  \ref{tbl1}, our sample set included 11 tablet types (A, B, ..., M), and for each tablet type there were three replicate samples (A-1, A-2, A-3, B-1, ...) to account for the variations in each tablet type, resulting in 33 samples. All tablets were flat-faced, with a nominal weight of 300 mg, a diameter of 10.0 mm, and a thickness, $l$, in the range of 2.97–3.25 mm, measured using a digital micrometer. The tablet density was calculated from the tablets dimensions and weight.

\begin{table*}[width=1.3\linewidth,cols=5,pos=h]
\caption{\textrm {Design of experiments for the tablets, a fraction of a full factorial design from reference \citep{Sparen2015MatrixSpectroscopy}. Design factors are filler particle size at two levels (91, 450 $\mu$m) and API concentration at five levels (16, 18, 20, 22, 25 w/w\% ). There are three replicates for each tablet type. } }

\begin{tabular*}{\tblwidth}{@{} ccccc@{} }
\toprule
Tablet & API particle & Filler particle & API concentration & Tablet density\\
 &  size d$_{50}$ ($\mu$m)  & size d$_{50}$ ($\mu$m)& (w/w\%)& (g cm$^{-3})$\\
\midrule
Type A\\
 A-1 & 71 & 91 & 16.8 & 0.3061 \\
A-2 & 71 & 91 & 16.8 & 0.3059 \\
A-3  & 71 & 91 & 16.8 & 0.3058 \\
  
Type B\\
B-1 & 71 & 450 & 16.0& 0.3248 \\
B-2 & 71 & 450 & 16.0& 0.3248 \\
B-3 & 71 & 450 & 16.0& 0.3246 \\

Type C\\
 C-1 & 71 & 91 & 18.5 & 0.3033\\
 C-2 & 71 & 91 & 18.5 & 0.3031\\
 C-3 & 71 & 91 & 18.5 & 0.3042\\
  
Type D\\  
D-1 & 71 & 450 & 18.4 & 0.3206\\
 D-2 & 71 & 450 & 18.4 & 0.3198\\
 D-3 & 71 & 450 & 18.4 & 0.3220\\ 
 
 Type E\\ 
E-1 & 71 & 91 & 20.6 & 0.3043 \\
E-2 & 71 & 91 & 20.6 & 0.3043 \\
E-3  & 71 & 91 & 20.6 & 0.3042 \\
 
 Type F\\ 
F-1 & 71 & 450 & 19.5 & 0.3219\\
F-2 & 71 & 450 & 19.5 & 0.3209\\
F-3 & 71 & 450 & 19.5 & 0.3210\\
  
Type G\\
 G-1& 71 & 91 & 22.7& 0.3014 \\
  G-2& 71 & 91 & 22.7& 0.3014 \\
   G-3& 71 & 91 & 22.7& 0.3021 \\
  
Type H\\  
  H-1& 71 & 450 & 22.9 & 0.3200 \\
 H-2& 71 & 450 & 22.9 & 0.3207 \\
  H-3& 71 & 450 & 22.9 & 0.3191 \\

Type K \\ 
 K-1& 71 & 91 & 24.8& 0.3025\\
 K-2& 71 & 91 & 24.8& 0.3014\\
  K-3& 71 & 91 & 24.8& 0.3023\\

Type L\\
 L-1& 71 & 450 & 25.0& 0.3180 \\
 L-2& 71 & 450 & 25.0& 0.3167 \\
  L-3& 71 & 450 & 25.0& 0.3186\\
  
Type M\\  
 M-1& 95 & 211 & 20.1 & 0.3010\\
 M-2& 95 & 211 & 20.1 & 0.3009\\
  M-3& 95 & 211 & 20.1 & 0.3003\\
\bottomrule
\end{tabular*}
  \label{tbl1}
 
\end{table*}

\begin{figure}[h!] 
    \centering
        \includegraphics[scale=0.17]{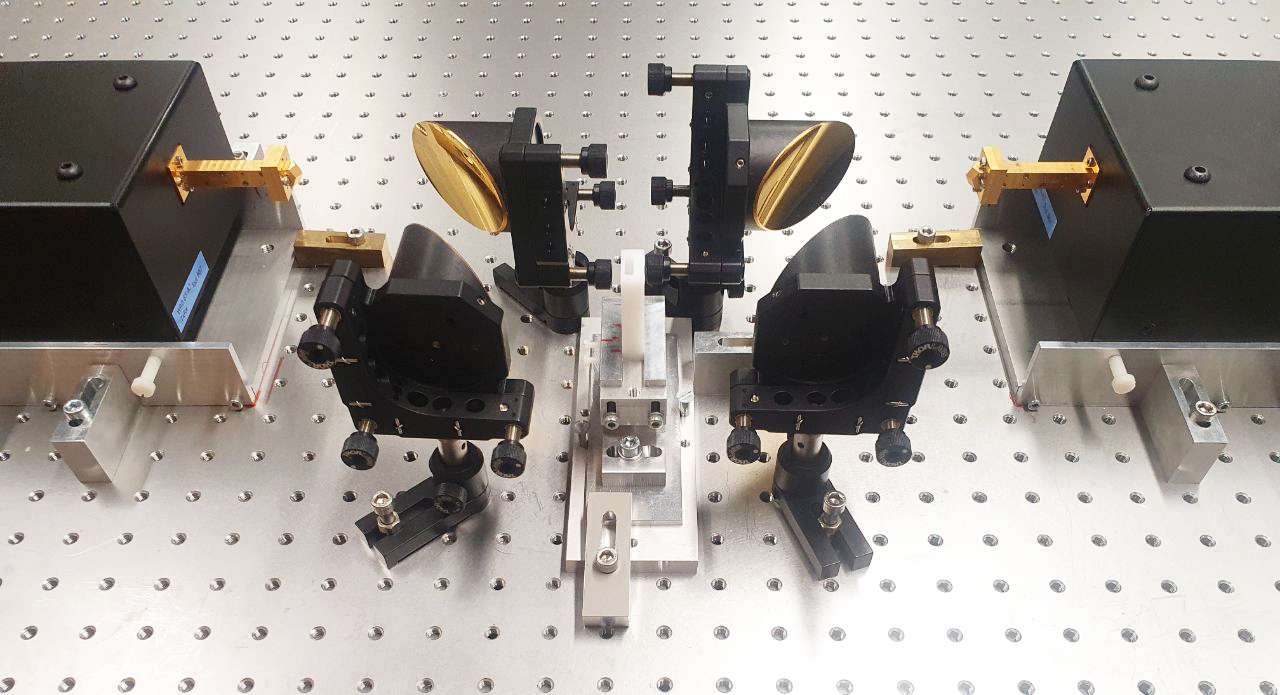}
    \caption{\textrm{ A photograph of the experimental set-up showing THz S-parameter measurements in transmission mode.}}
     \label{fig: 1}
\end{figure}

\subsection{ Measurement set-up}

To obtain the attenuation spectra of the tablets, the complex transmission coefficients were measured \textcolor{black}{in frequency domain from 750 GHz to 1.5 THz, with the frequency step of 0.925 GHz}. The measurement set-up consisted of a vector network analyzer (VNA) (Keysight PNA-X), frequency extenders WM380 (750 GHz-1.1 THz) and WM250 (1.1-1.5 THz), horn antennas, and four off-axis parabolic mirrors, see Fig.~\ref{fig: 1}. The frequency range was selected based on the presence of the absorption peaks of Mannitol and Ibuprofen in the spectra. During measurements, the intermediate frequency bandwidth was set to 10 Hz in order to minimize the measurement noise level. The signal was focused on the center of the tablets using the parabolic mirrors with a focal length of 76.4 mm. The estimated mid-band beam diameter at the focal point is approximately 1.3 mm at 750 GHz to 1.1 THz and 0.9 mm at 1.1-1.5 THz. Each sample was measured twice by dismounting and mounting the tablets in the sample holder to account for the measurement errors like the white noise of the VNA and tablet displacement in the sample holder. This led to 66 measurements ranging from 750 GHz to 1.5 THz. Before each tablet measurement, the empty sample holder was measured as a reference for relative measurements, which was the ratio between the transmitted signal in the presence of the tablets versus the empty holder. \textcolor{black}{The measurement environment could have been purged with dry nitrogen gas to remove the water vapour and to increase the SNR. In our case, we decided not to use this procedure in order to mimic the production line conditions.}

\subsection{ Data processing } \label{Sec 2.2}
The attenuation coefficient of the samples, $\alpha (f)$, was calculated from the measurement of the scattering parameters using equation \ref{eq: 1}~\citep{Chen2017IsomersSpectroscopy}:
\begin{equation}
    \alpha (f)=-\frac{1}{l}log\left|\frac{T_s}{T_a}\right|^2, \label{eq: 1}
\end{equation}
where $l$ is sample thickness in $cm$, and $T_s$ and $T_a$ are the measured complex transmission coefficients ($S_{21}$) of the sample and empty sample holder (air), respectively, presented in supplementary data.

Noise, high-frequency oscillations, and a slanted baseline were present in the measured THz spectra. The oscillations were caused by standing waves in the set-up, and the slanted baseline was caused by scattering effects and frequency-dependent absorption. In order to reduce the effect of such phenomena, the spectra were pre-processed using a combination of smoothing, baseline correction, and normalization filters. \textcolor{black}{It should be noted that in the case of amorphous materials, the baseline contains information about the content and should not be removed.} The data pre-processing was performed in SIMCA 17 (Sartorius Stedim Data Analytics, Umeå, Sweden), which is a software for multivariate data analysis.

In the next step, the treated spectra were used to create a model for predicting the API concentration and tablet density, using multivariate analysis. 
The measured samples were divided into two sets, including a training set and a test set. The training set was used to build the model, and the test set was used to evaluate the robustness of the model for unknown samples. From the three replicates of each tablet type, the first two samples were used in the training set, and the third sample was used in the test set. This led to 44 samples in the training set and 22 samples in the test. The samples were labeled such that the first letter shows tablet type, the second part shows the number of the replicate sample, and the third part shows the number of measurements for each sample.  

\subsection{Multivariate analysis}
Multivariate analysis is an important tool for qualitative and quantitative analysis of multivariate spectral data. We used orthogonal partial least squares (OPLS) regression~\citep{Trygg2002OrthogonalO-PLS}  to establish a prediction model for the quantification of API concentration and tablet density.  \textcolor{black}{OPLS is a modification of the traditional partial least squares (PLS) regression. OPLS separates the variability in the X spectral data into two parts, one that is predictive of the Y response and another that is orthogonal to Y. In the case of just a single Y variable and with the condition of the same number of components, both PLS and OPLS models fitted to the same data will give identical predictions.} In the calibration set, the variances explained for the calibration set ($R^2$) and the cross-validation ($Q^2$), the root-mean-square of the calibration (RMSEC) and the cross-validation (RMSECV) were used to assess the model performance. Cross-validation is a model validation method that uses different portions of the calibration set iteratively to train and test the model. In order to further challenge the robustness of the model for the prediction of unknown samples, a test set was applied to the model. The root-mean-square error of prediction (RMSEP) and mean bias error for the prediction (MBEP) were used to assess how well the test samples fit the calibration model. Relative prediction error (RPE) is the RMSEP normalized with the mean value of the response, measuring the accuracy of the prediction regardless of the nature of the response.

\section{Results and discussions}
\begin{figure}[b] 
 \centering
        \includegraphics[scale=1]{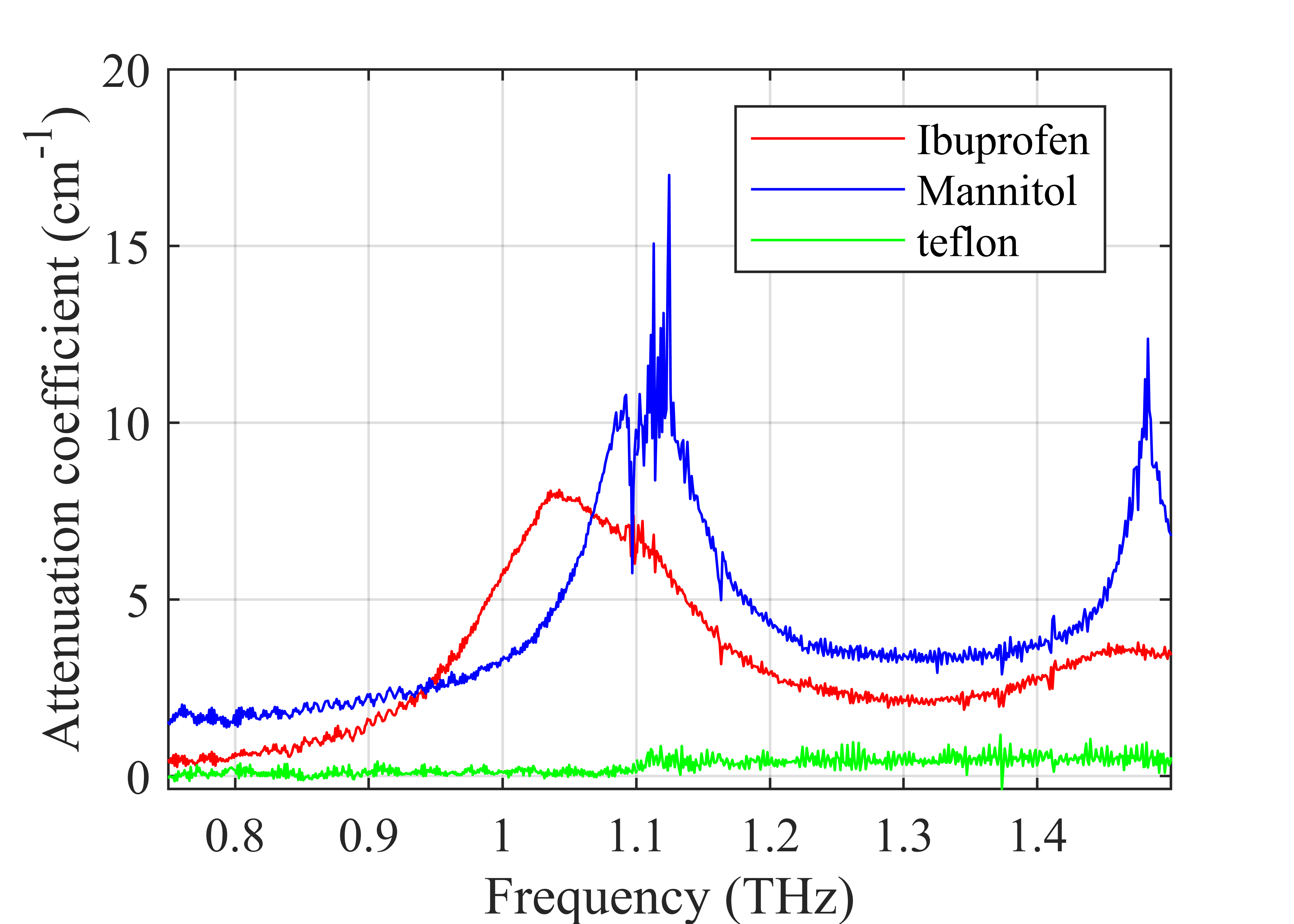}
       \small {(a)} 
        \includegraphics[scale=1]{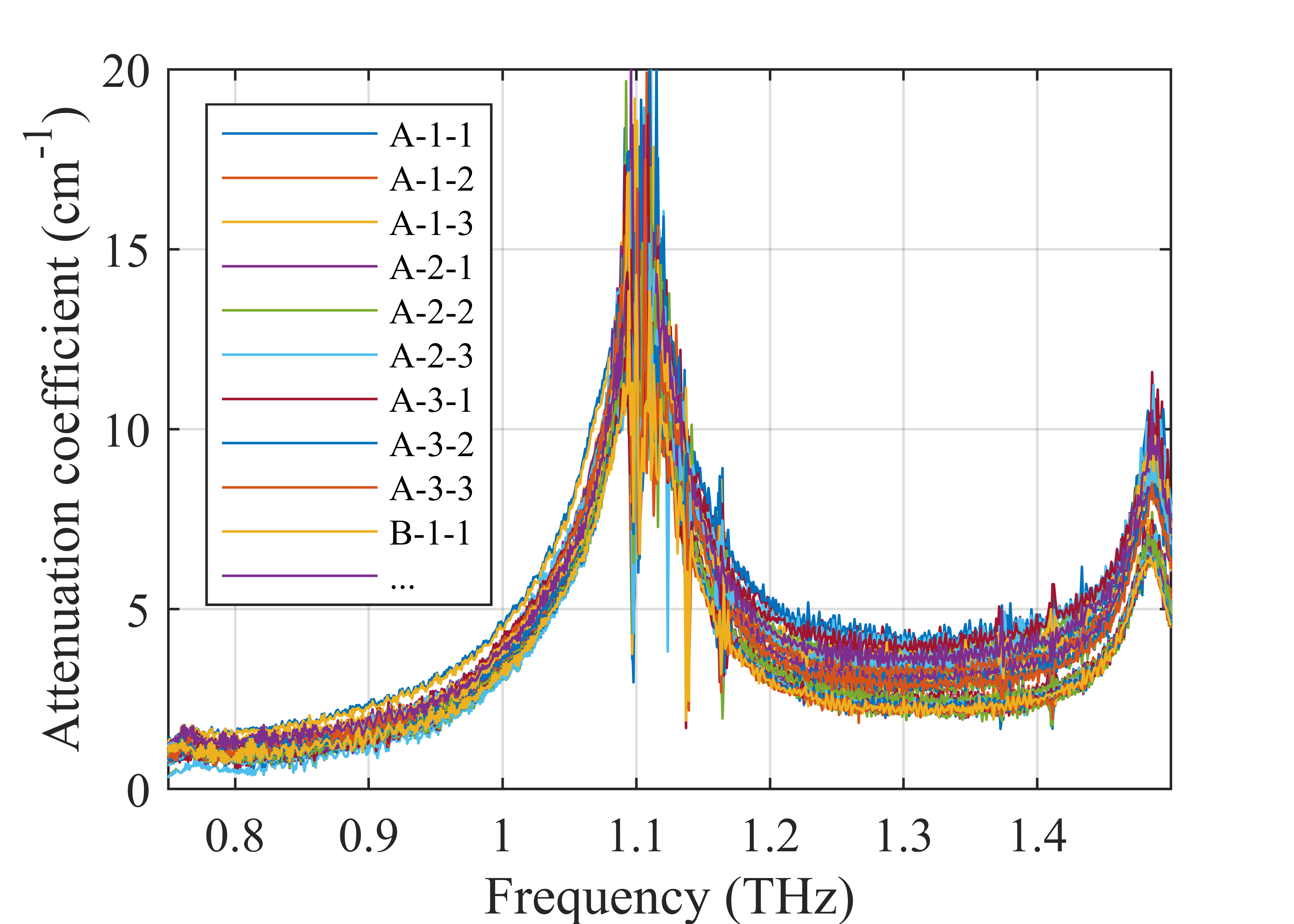}
        \small {(b)}
    \caption{\textrm{ Spectral data for a) Ibuprofen, Mannitol, and Teflon as a reference. b) samples from table \ref{tbl1}}.}
     \label{fig: 2}
\end{figure}
One tablet consisting of pure ibuprofen and one tablet of pure Mannitol were measured to identify the absorption peaks of each tablet ingredient individually across the frequency range.
Fig. \ref{fig: 2}a shows the attenuation spectra of pure Ibuprofen, pure Mannitol, and Teflon. Ibuprofen showed two broad peaks centered around 1.05 THz and 1.47 THz. Mannitol showed two peaks centered around 1.10 THz and 1.48 THz, in agreement with Mannitol type $\beta$~\citep{Allard2011ImprovedMeasurementsb}.  The peaks occurring at around 1.10 THz were not completely resolved due to the lower signal-to-noise ratio (SNR) at the beginning of the frequency band of the extenders, the low SNR at absorption peaks, and water absorption lines. Teflon was just used for reference and as expected, did not show any absorption peaks. Fig. \ref{fig: 2}b shows the attenuation spectra of all measured tablets, and like Mannitol, the peaks at around 1.10 THz were not completely resolved (see supplementary material). \textcolor{black}{In the THz spectra of the samples, the peak seen at around 1.1 THz and below 1.5 THz are the superposition of the Ibuprofen and Mannitol peaks that heavily overlap. Therefore, simple univariate calibration method would not be selective for this formulation.} 

\subsection{OPLS model for API concentration}
\begin{figure}[h]
    \centering
    \includegraphics[scale=0.85]{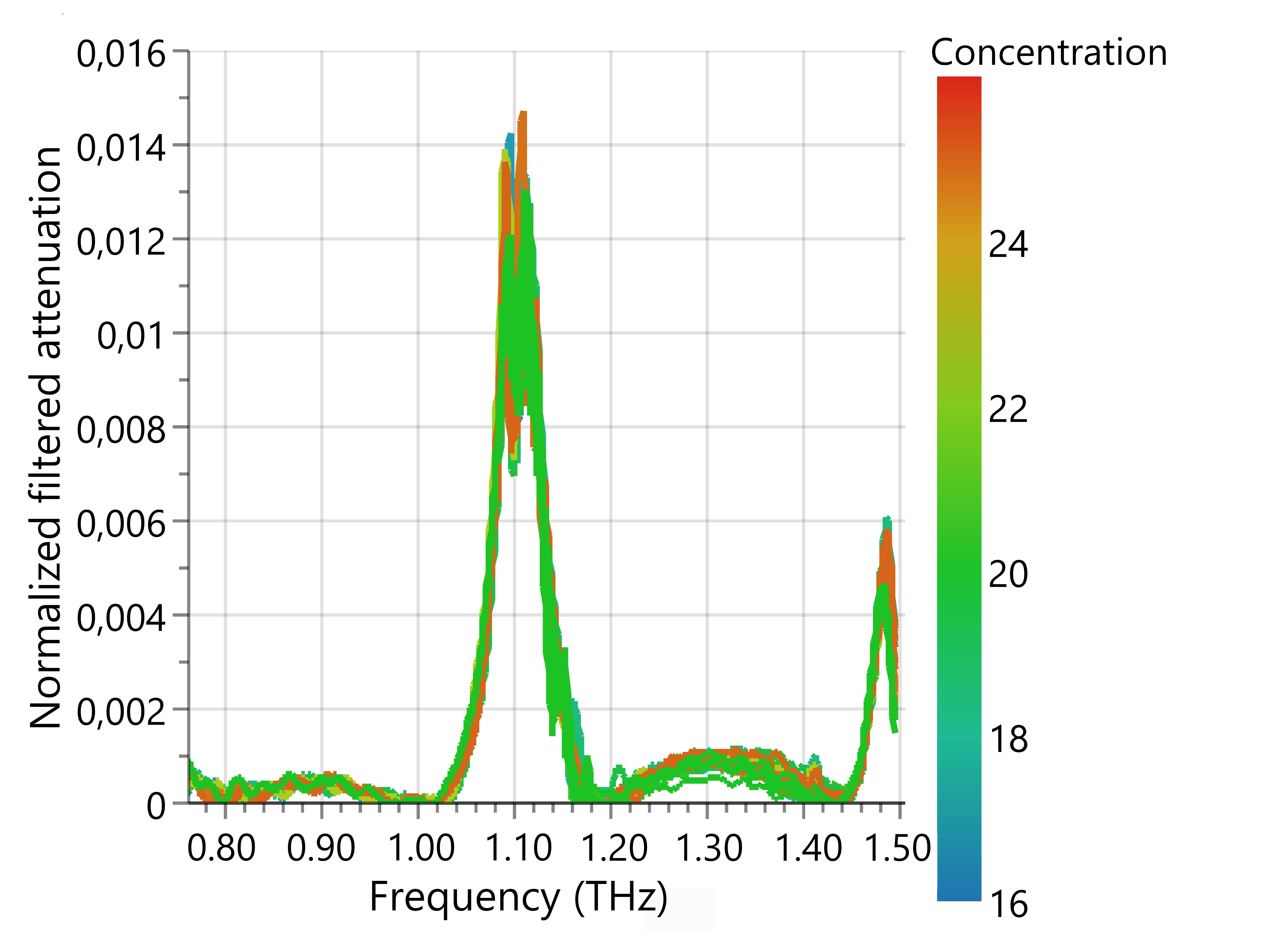}
    \caption{Pre-processed spectral data for API concentration, after noise effect reduction, baseline correction, and normalization. }
    \label{fig: 3}
\end{figure}

To model the API concentration, first the attenuation data were pre-processed, and then an OPLS analysis was applied to the data to obtain a prediction model. For the pre-processing, a Savitzky-Golay smoothing filter~\citep{Savitzky1964SmoothingProcedures} was used to reduce the effect of noise and oscillations. \textcolor{black}{The filter was calculated from moving quadratic sub-models with 31 data points long, and edge effects were excluded.} Then, an asymmetric least squares algorithm~\citep{Peng2010AsymmetricCorrection} was used for each spectrum to correct baseline variations. Finally, each spectral value was normalized with respect to the sum of the signal intensity over the frequency range. The treated attenuation spectra of the tablets are shown in Fig. \ref{fig: 3}, where different API concentrations are marked with different colours. 

In the next step, OPLS regression was applied to the scaled and centered training set to establish a prediction model for API concentration, having the frequency points as the $X$ independent variables and API concentration as the $Y$ response variable. \textcolor{black}{For the characterisation of API, the parts of the spectrum including the absorption peaks have the dominant role in modeling.
A first data evaluation revealed differences in the THz spectra of the tablets measured on a second day under higher humidity conditions, which complicated the modelling. After excluding those measurements affected by humidity variation (tablets B and H), we succeeded
in developing a good model with $R^2$ and $Q^2$ values of $0.98$ and $0.92$, respectively. The training set consisted of 36 samples and the test set of 18 samples.} 


\begin{figure}[h]
    \centering
    \includegraphics[scale=0.85]{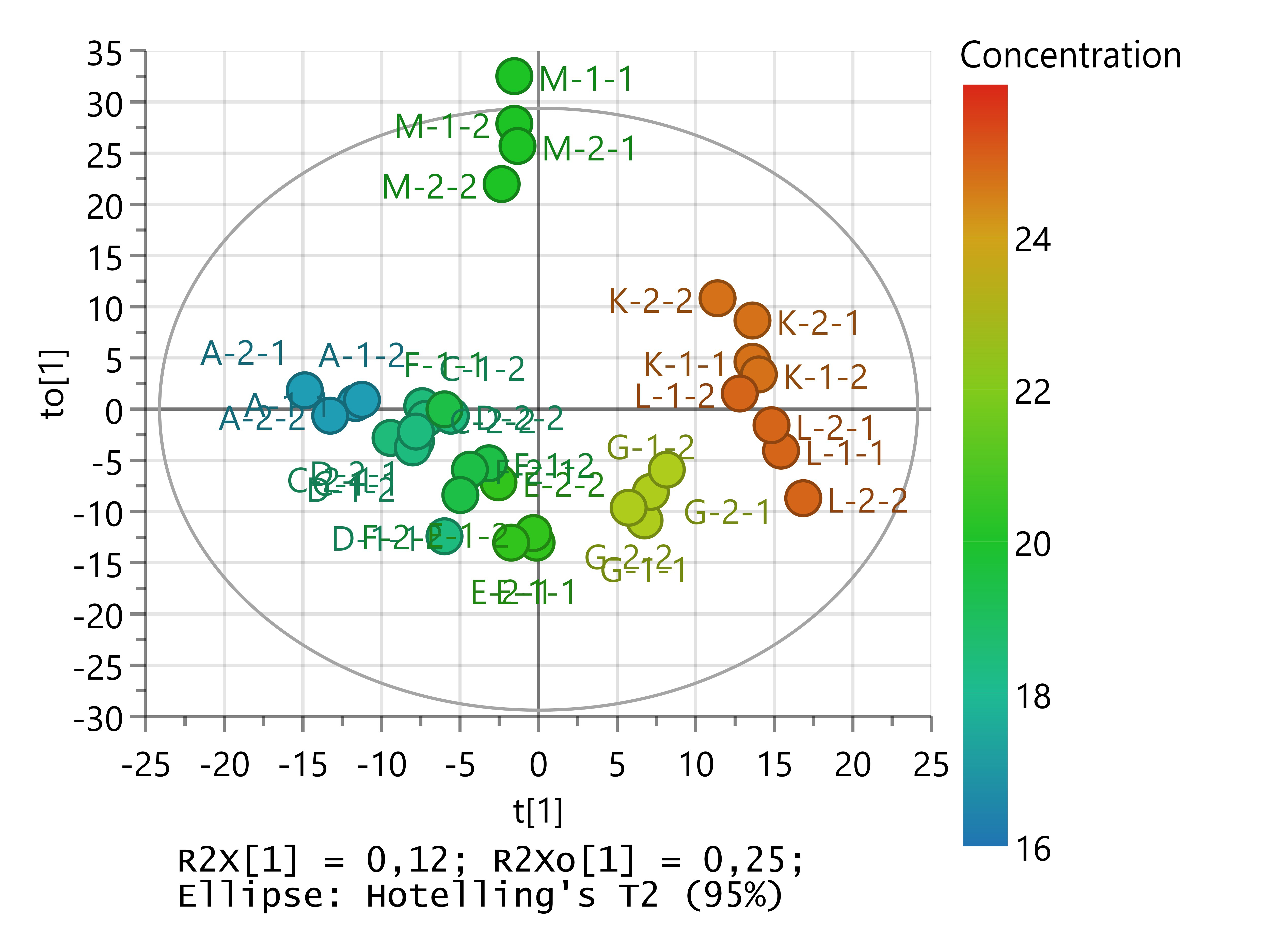}
    \caption{ Score plot for API concentration. Tablets are clustered with respect to the API concentration across the horizontal axes. Tablets are labeled with three digits that represent type, replicate, and measurement number, respectively.  }
    \label{fig: 4}
\end{figure}

\begin{figure}[h]
    \centering
      \includegraphics[scale=0.85]{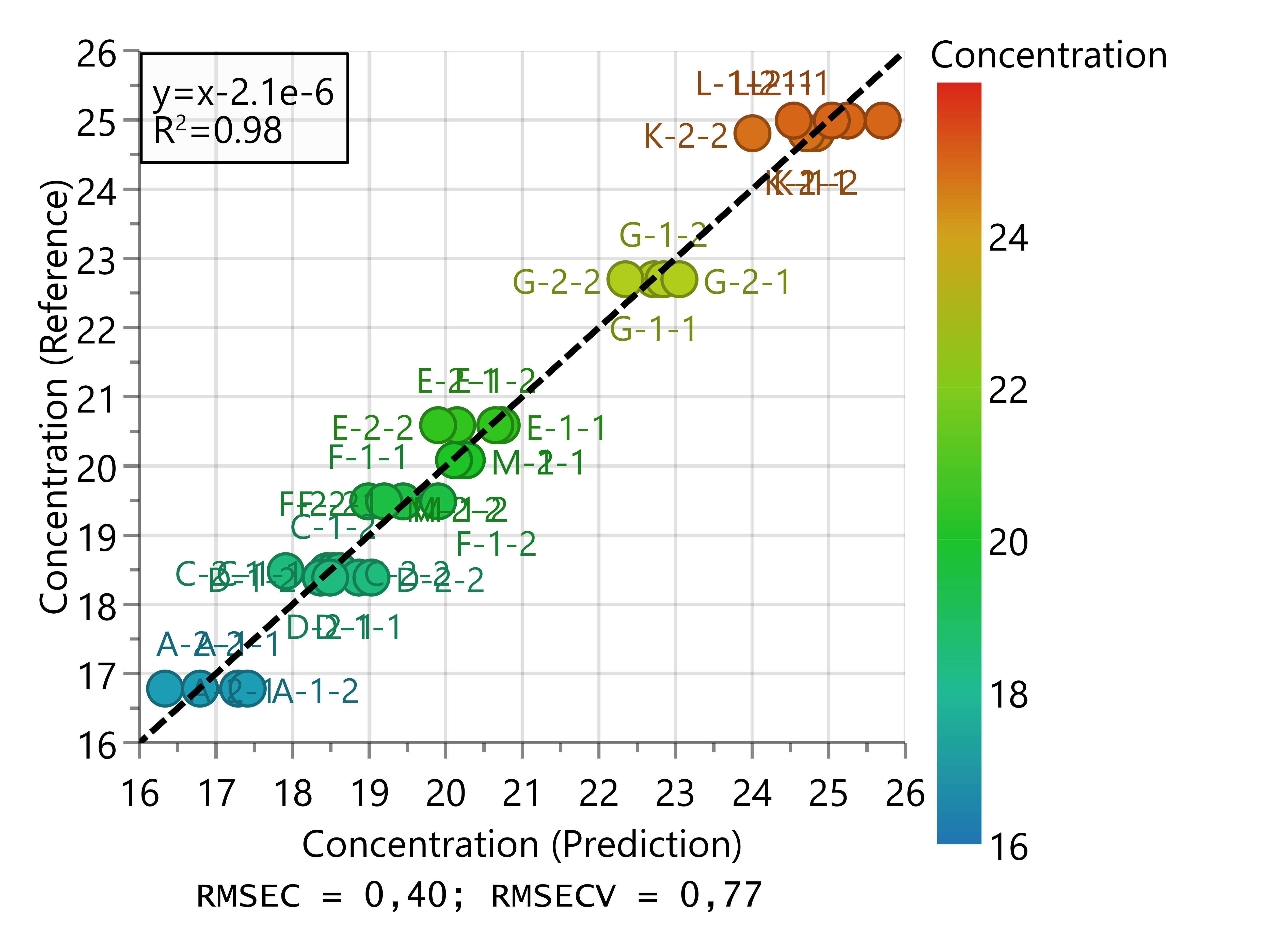}
    \small {(a)}
    \includegraphics[scale=0.85]{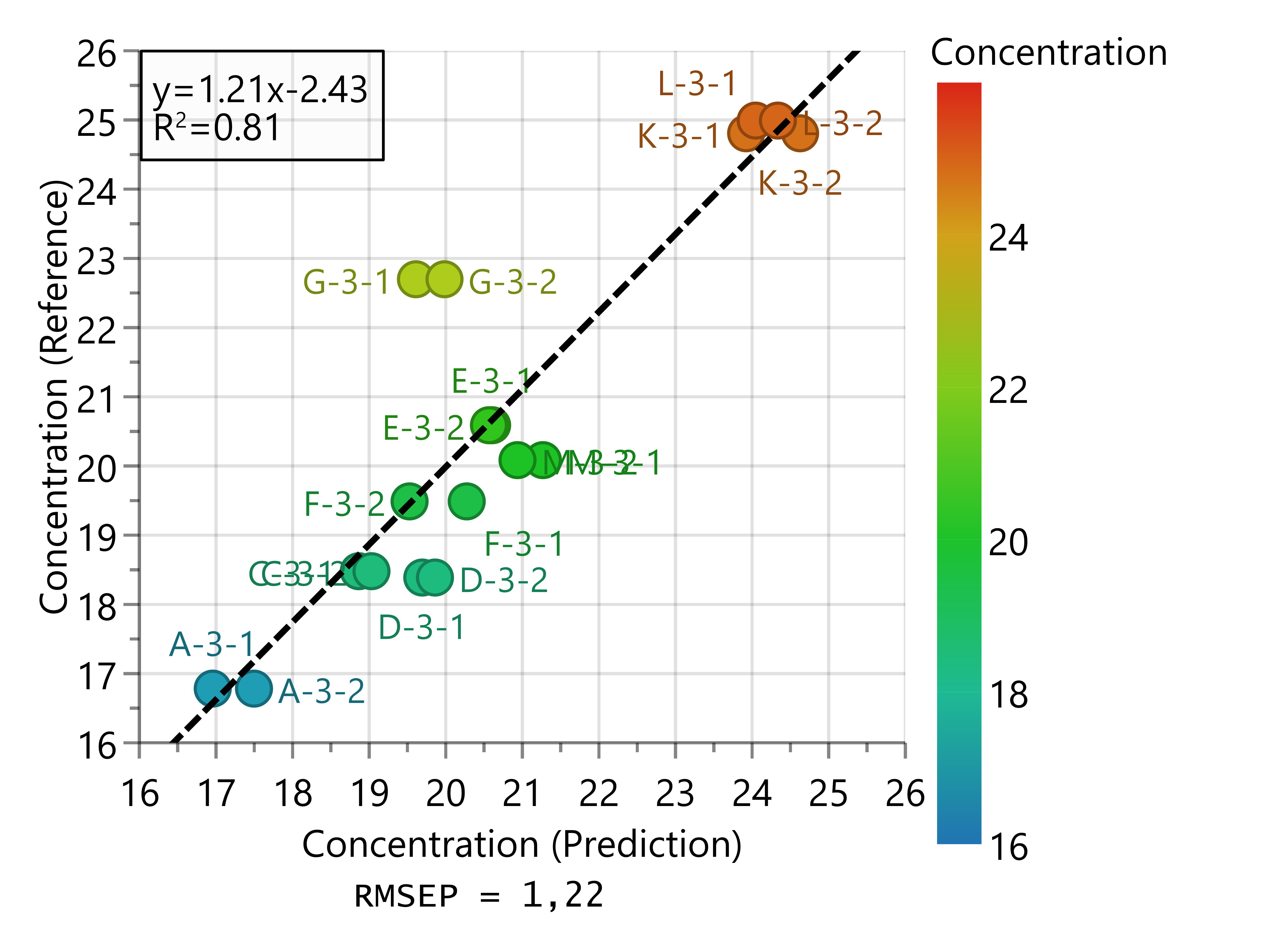}
    \small {(b)}  
    \caption{Prediction plot for API concentration a) calibration set b) test set. Tablets are labeled with three digits that represent type, replicate, and measurement number, respectively.}
    \label{fig: 5}
\end{figure}

An OPLS score plot in Fig. \ref{fig: 4} shows the clusters of the training samples based on their similarity for modeling the API concentration. The variations across the horizontal axes show the predictive component for the response, and the variations across the vertical axes show the orthogonal components, which are unrelated to the response and express the variations within clusters.
In this plot, the  training samples are clustered across the horizontal direction corresponding to the five API levels. Tablets are ordered from left to right by increasing API concentration. Moreover, the variations between replicate samples for each tablet type (tablets labeled with the same first digit but different second digit) and the repeatability of the measurements (different third digit in the label) can be observed in this plot.

Fig. \ref{fig: 5}a shows the observed versus predicted values for API concentration based on the OPLS regression. The $R^2$ was 0.98, and the RMSEC and RMSECV were 0.40 and 0.77, respectively. Five clusters of samples are observed corresponding to the five levels of API concentration. 
To challenge the robustness of the OPLS model, a test set was used. Fig. \ref{fig: 5}b shows the comparison between the reference and prediction results for the test samples with five clusters corresponding to the API concentration. \textcolor{black}{For Sample G-3, we obtained predicted values about 3\% lower than the expected API level. One possible reason for this could be that the API may not have been evenly distributed across the tablet, while  the THz beam illuminated mainly the center of the samples, which could lead to sub-sampling. The terahertz spectrum for this sample indicated a lower level of API than expected (data not shown), which supports this hypothesis.} The relative prediction error and RMSEP were 6\% and 1.22, respectively, which were less good than the results in paper \citep{Sparen2015MatrixSpectroscopy} because Raman and NIR techniques spectra have more selective bands and contain more chemical information than THz spectroscopy. \textcolor{black}{It must be noticed that in \citep{Sparen2015MatrixSpectroscopy}, a larger number of tablets were included in the model.} The quantification of the API concentration could probably be further improved by measuring \textcolor{black}{in a controlled environment} and completely resolving the absorption peaks. \textcolor{black}{The latter} can be achieved by purging the measurement system to remove the water lines and/or averaging the measurements over time, with the cost of increasing the complexity of the system and the measurement time. 

\subsection{OPLS model for tablet density}
\begin{figure}[t]
    \centering
    \includegraphics[scale=0.8]{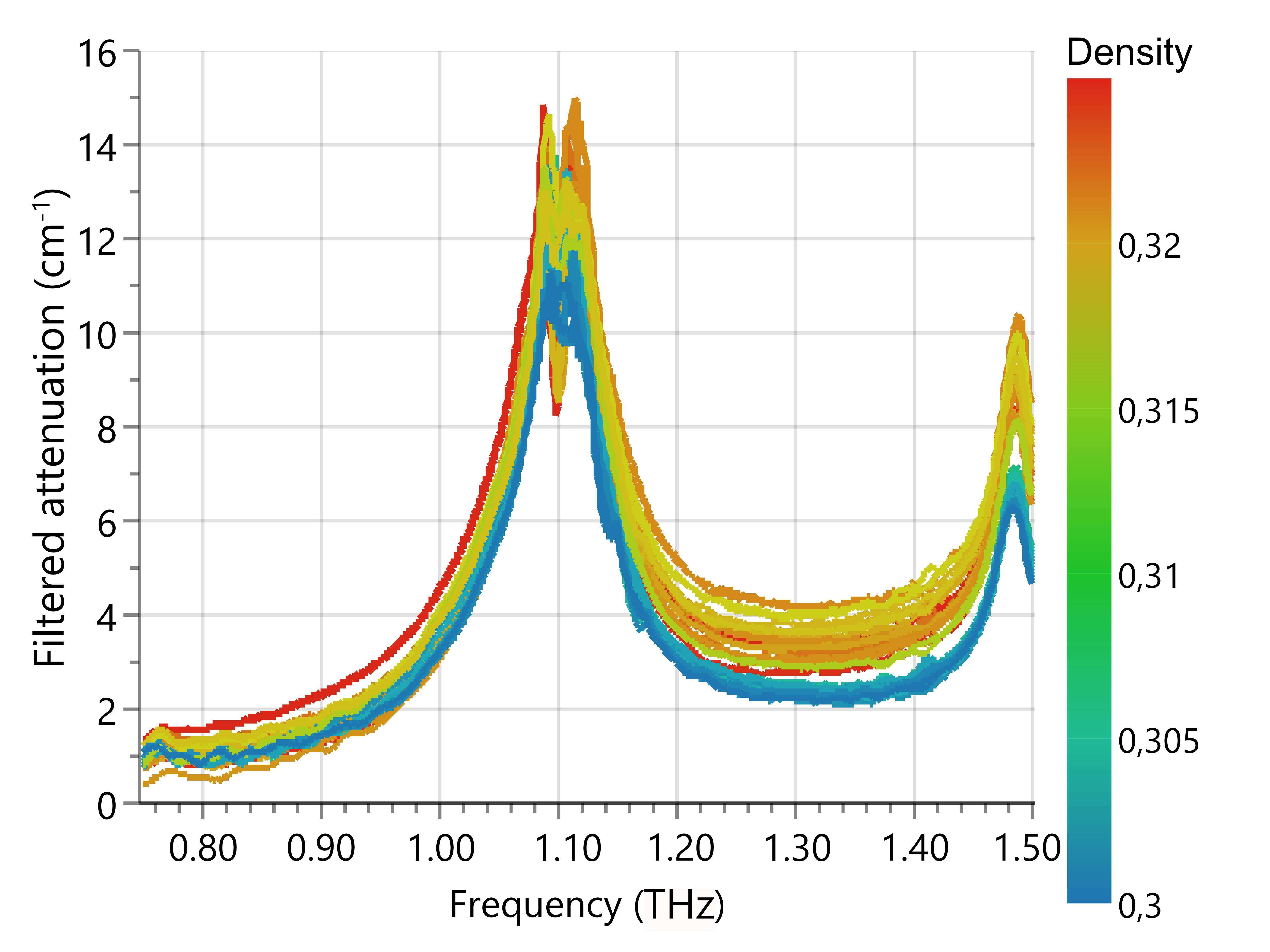}
    \caption{Pre-processed spectral data for the quantification of tablet density after applying Savitzky-Golay smoothing filter for removing the high frequency noise.}
    \label{fig: 6}
\end{figure}

To model the tablet density, the spectral data were treated to reduce the noise and oscillations using the Savitzky-Golay smoothing filter. In this case, the baseline correction and data normalization were not conducted because of the correlation between the baseline and tablet density through the particle size and porosity~\citep{Moradikouchi2022TerahertzTablets}. 
Fig. \ref{fig: 6} shows the treated spectra after applying the smoothing filter, where different tablet densities were assigned with a different color. As can be observed, the spectral data were clustered into two groups, corresponding to the tablets with different initial particle sizes. The attenuation for tablets with smaller initial particle sizes was lower than those with larger particles. At the end of the band, tablets with larger particle size showed higher attenuation due to scattering effects~\citep{Shen2008EliminationSpectroscopy, Wu2008ProcessApplication}.

\begin{figure}[h]
    \centering
    \includegraphics[scale=0.7]{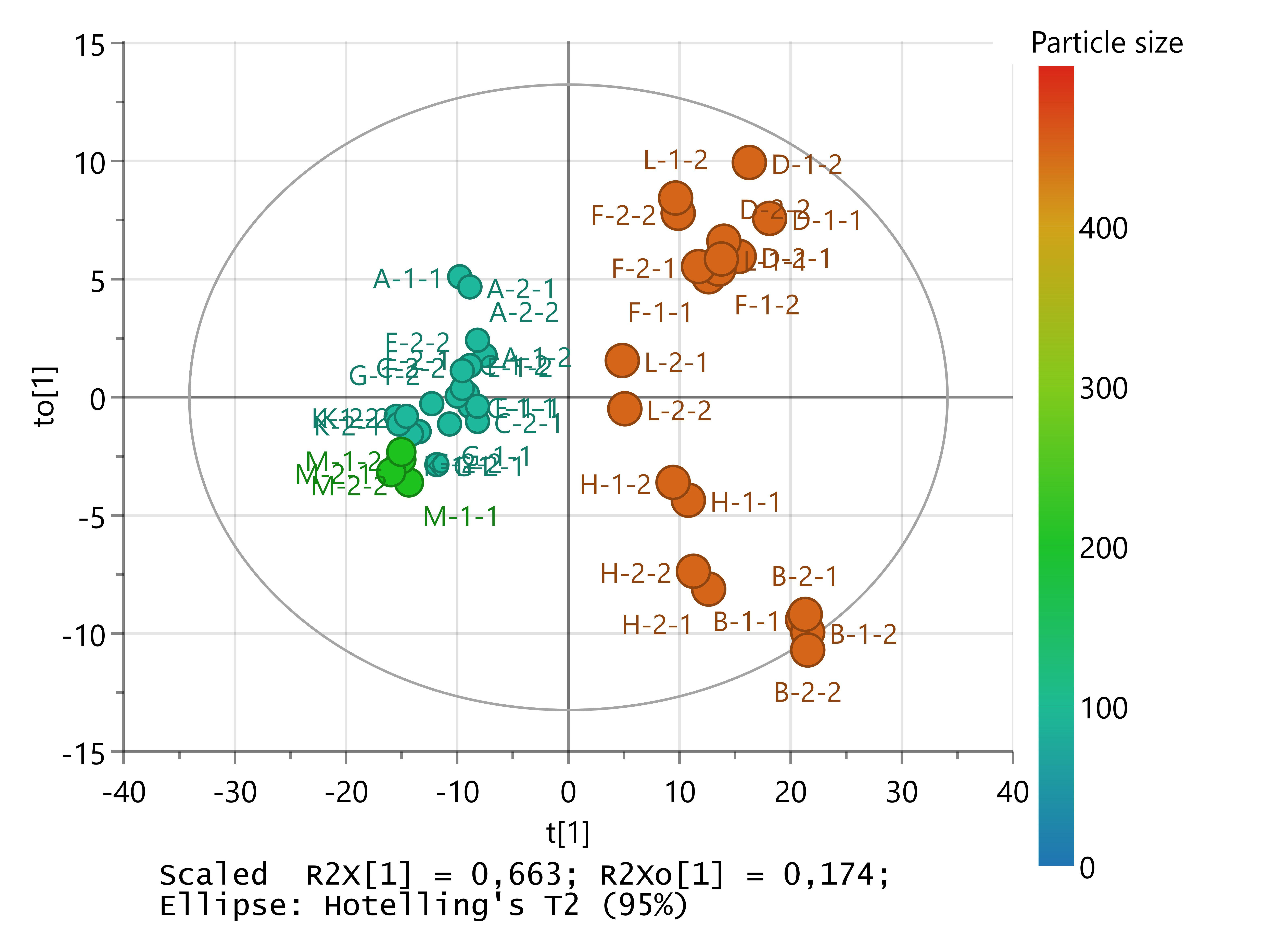}
    \caption{Score plot for tablet density. Tablets are clustered across the horizontal axes with respect to particle size. Tablets are labeled with three digits that represent type, replicate, and measurement number, respectively. Small and large particle sizes are shown with small and large circles, respectively.}
    \label{fig: 7}
\end{figure}

Next, OPLS regression was applied to the 44 training samples, having the spectral data centered, to establish a prediction model for the tablet density.  \textcolor{black}{For the characterization of the
tablet density based on the attenuation, higher frequencies play the dominant role due to the correlation between tablet density and particle size.}
The OPLS score plot in Fig. \ref{fig: 7} shows two clusters of training samples across the horizontal direction with respect to the particle size, expressing the correlation between tablet density and the variations in the particle size. Tablets with larger particles are clustered on the right side marked with larger circles, and tablets with smaller particle sizes are on the left side marked with smaller circles. 
The vertical direction expresses the variation within the clusters due to API concentration. 

\begin{figure}
    \centering
         \includegraphics[scale=0.7]{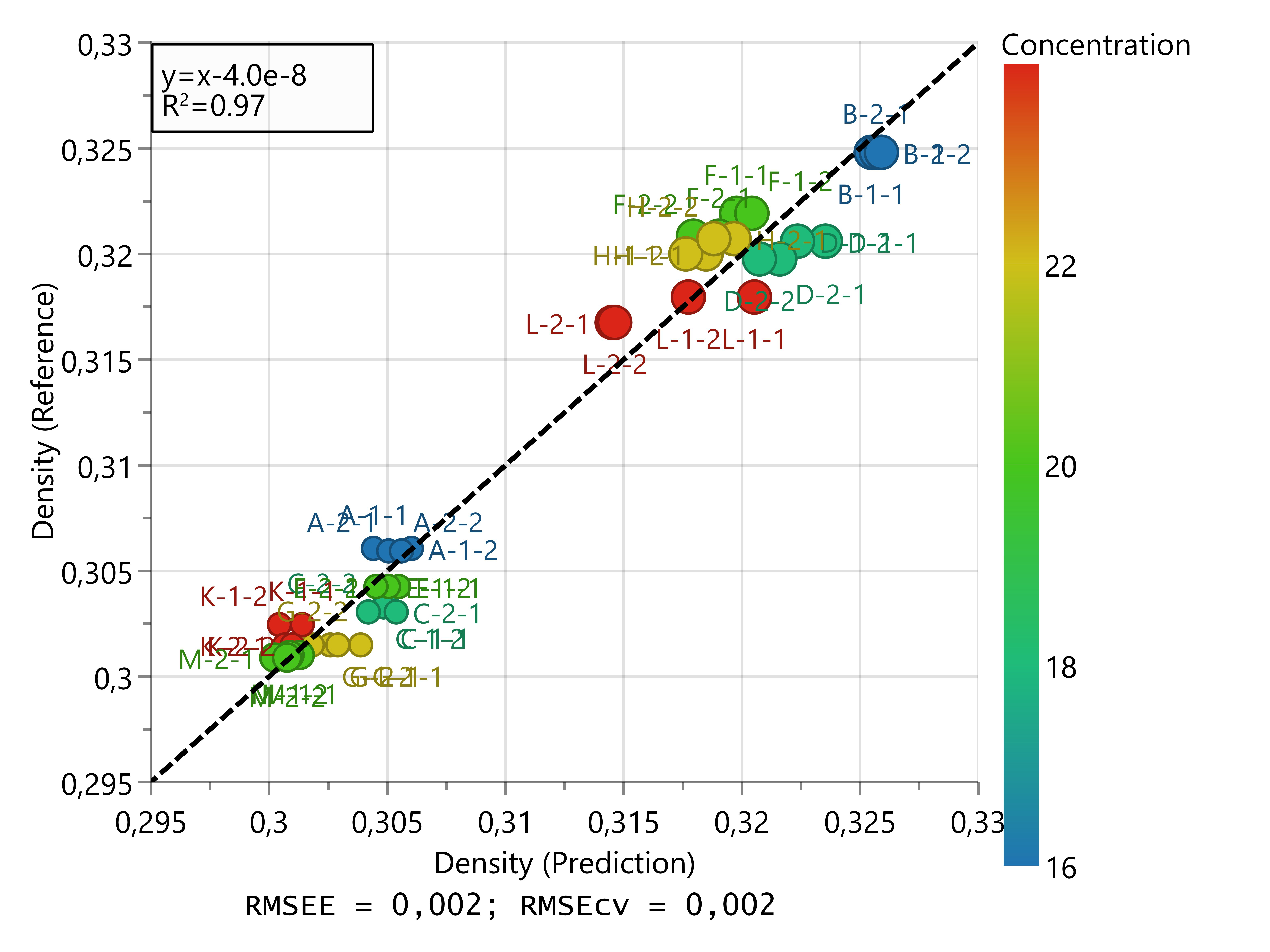}
         \small {(a)} 
         \includegraphics[scale=0.7]{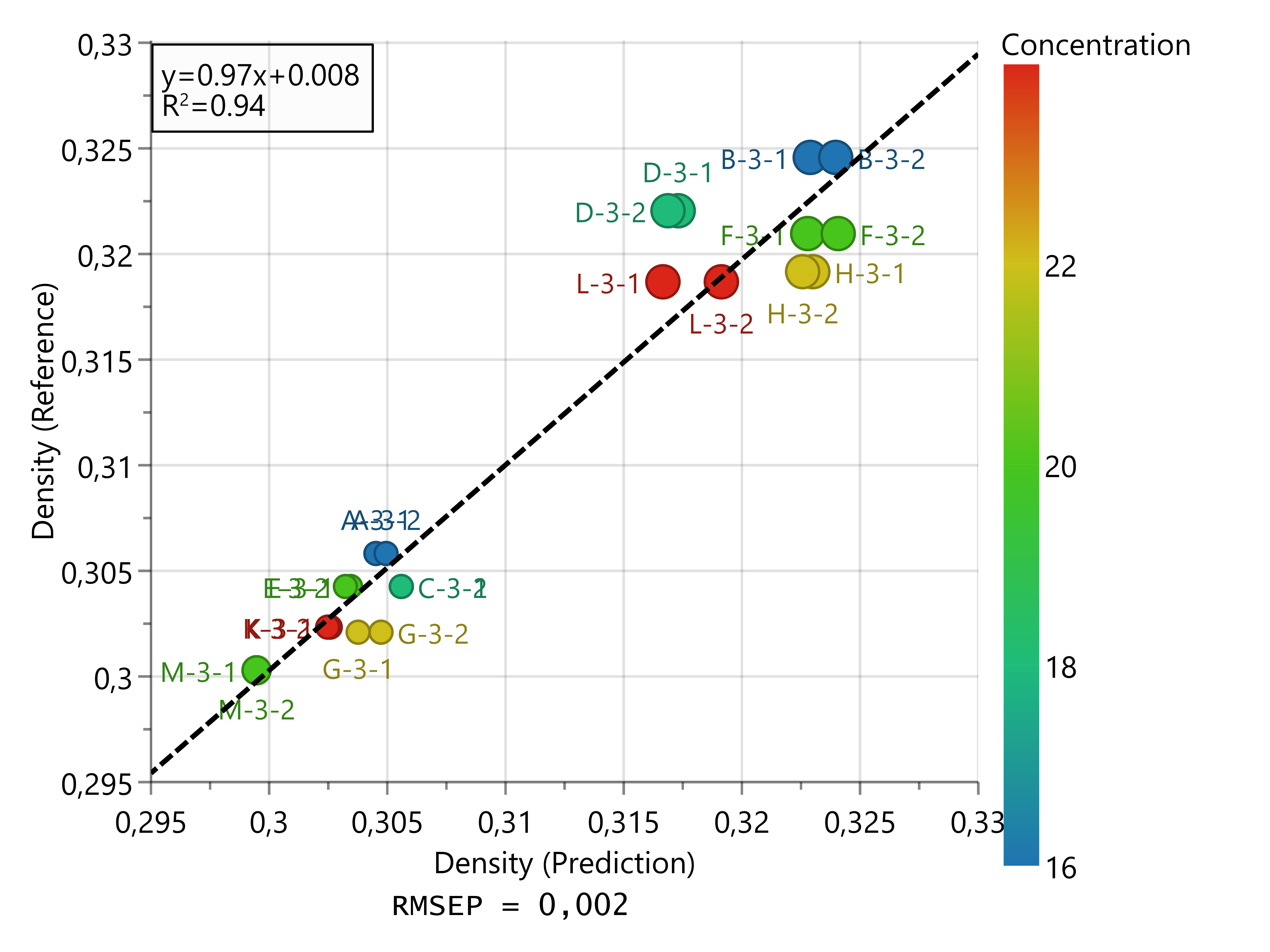}
         \small {(b)} 
    \caption{Prediction plot for tablet density a) calibration set b) test set. Tablets are labeled with three digits that represent the type, replicate, and measurement number, respectively. Small and large particle sizes are shown with small and large circles, respectively.}
    \label{fig: 8}
\end{figure}

Fig. \ref{fig: 8}a shows the reference versus prediction values for tablet density.  The RMSEC and RMSECV for the obtained OPLS model were both 0.002, and the $R^2$  was 0.97. These values show the high accuracy and the goodness of the regression model. \textcolor{black}{ It should be noticed that for the quantification of tablet density, samples measured under higher humidity condition (B and H)  were not excluded, indicating that humidity variation in the lab environment was not as critical as for quantifying the API concentration.}

The robustness of the obtained OPLS model was further assessed by applying the 22 test samples to the model. Fig. \ref{fig: 8}b presents the comparison between the reference and prediction results for the test samples. The prediction of tablet density on the test samples gave a relative prediction error of  0.7\%, and RMSEP and MBEP of 0.002 and 0.0002, respectively. These results show that the OPLS model for tablet density is significantly accurate and precise and performs better than the OPLS  model for API concentration. The summary of the two models' performance is given in table \ref{tab: 2}. Moreover, in both Figs. \ref{fig: 8} a and b, we see two clusters of tablets corresponding to the small and large particle sizes marked with smaller
and larger circles respectively, showing that initial particle size  impacted the tablet density. Besides that, the results show that within each cluster of tablets, as the API concentration increased, the tablet density decreased being in line with the results from our previous study~\citep{Moradikouchi2022TerahertzTablets}. Also, it can be observed that initial particle size has a higher impact than API concentration on the tablet density. 

\begin{table*}[cols=6,pos=h]

    \caption{A summary of the OPLS models and their performance for the quantification of API concentration and tablet density. }

    \begin{tabular}{ccccccccc}
        \toprule
     Model & Calibration set  & Test set& $R^2$ & $Q^2$&RMSEC& RMSEP  & MBEP & RPE  \\
           &  &  &  &    && (test set) & (test set) &  (test set) \\
        \midrule
       
         API concentration & 36 samples&  18 samples & 0.98 & 0.92 &0.40& 1.22 & -0.06 & 6.0\%\\
         Tablet density & 44 samples & 22 samples  &0.97 & 0.94 &0.002& 0.002 & -0.0002 & 0.7\% \\
         \bottomrule
    \end{tabular}
    
    \label{tab: 2}
\end{table*}

\section{Conclusion}
In this paper, we explored the THz-FDS technique together with multivariate data analysis for the characterization of pharmaceutical tablets. A quantitative OPLS method was proposed to model the API concentration and tablet density based on terahertz transmission measurements at a frequency range of 750 GHz to 1.5 THz. The OPLS model for the API concentration achieved an $R^2$ value of 0.98 and a relative prediction error of 6\% for tablets with a 2\% difference in the level of the API concentration. The results showed that THz-FDS is less precise than NIR and Raman spectroscopy for the quantification of API concentration, but it brings the advantage of being less affected by particle size variations, which could prevent systematic bias errors. \textcolor{black}{To optimize the determination of API content, the THz measurements should preferably be conducted in a controlled low humidity environment.}
The OPLS model for the tablet density achieved an $R^2$ value of 0.97 and an excellent relative prediction error of 0.7\%. This highlights the potential of terahertz spectroscopy for characterizing the physical properties of tablets, which is a challenge for other non-destructive techniques.
In summary, THz-FDS in combination with multivariate analysis showed to be a promising process analytical tool that complements NIR and Raman techniques for the characterization of the physical properties of tablets and API concentration with the advantage of low scattering effects.

\section*{Declaration of Competing Interest}
The authors declare that they have no known competing financial interests or personal relationships that could have appeared to influence the work reported in this paper.

\section*{Acknowledgment}
The measurements were carried out at the Kollberg Laboratory, at Chalmers
University of Technology, Gothenburg, Sweden.
\section*{Funding}
This research was funded by a grant from the Swedish foundation for strategic research (SSF), ID 17-0011, and from AstraZeneca, Gothenburg, Sweden.

\newcommand*{\doi}[1]{\href{https://doi.org/\detokenize{#1}}{ \detokenize{#1}}}
\bibliographystyle{cas-model2-names}
\bibliography{cas-refs}

\begin{thebibliography}{34}
\expandafter\ifx\csname natexlab\endcsname\relax\def\natexlab#1{#1}\fi
\providecommand{\url}[1]{\texttt{#1}}
\providecommand{\href}[2]{#2}
\providecommand{\path}[1]{#1}
\providecommand{\DOIprefix}{doi:}
\providecommand{\ArXivprefix}{arXiv:}
\providecommand{\URLprefix}{URL: }
\providecommand{\Pubmedprefix}{pmid:}
\providecommand{\doi}[1]{\href{http://dx.doi.org/#1}{\path{#1}}}
\providecommand{\Pubmed}[1]{\href{pmid:#1}{\path{#1}}}
\providecommand{\bibinfo}[2]{#2}
\ifx\xfnm\relax \def\xfnm[#1]{\unskip,\space#1}\fi
\bibitem[{Aina et~al.(2010)Aina, Hargreaves, Matousek and
  Burley}]{Aina2010TransmissionFormulations}
\bibinfo{author}{Aina, A.}, \bibinfo{author}{Hargreaves, M.D.},
  \bibinfo{author}{Matousek, P.}, \bibinfo{author}{Burley, J.C.},
  \bibinfo{year}{2010}.
\newblock \bibinfo{title}{{Transmission Raman spectroscopy as a tool for
  quantifying polymorphic content of pharmaceutical formulations}}.
\newblock \bibinfo{journal}{Analyst} \bibinfo{volume}{135}.
\newblock \DOIprefix\doi{10.1039/c0an00352b}.
\bibitem[{Allard et~al.(2011)Allard, Cornet, Debacq, Meurens, Houde and
  Morris}]{Allard2011ImprovedMeasurementsb}
\bibinfo{author}{Allard, J.F.}, \bibinfo{author}{Cornet, A.},
  \bibinfo{author}{Debacq, C.}, \bibinfo{author}{Meurens, M.},
  \bibinfo{author}{Houde, D.}, \bibinfo{author}{Morris, D.},
  \bibinfo{year}{2011}.
\newblock \bibinfo{title}{{Improved detection sensitivity of D-mannitol
  crystalline phase content using differential spectral phase shift terahertz
  spectroscopy measurements}}.
\newblock \bibinfo{journal}{Optics Express} \bibinfo{volume}{19},
  \bibinfo{pages}{4644}.
\newblock \DOIprefix\doi{10.1364/oe.19.004644}.
\bibitem[{Bawuah et~al.(2020)Bawuah, Markl, Farrell, Evans, Portieri, Anderson,
  Goodwin, Lucas and Zeitler}]{Bawuah2020Terahertz-BasedTutorial}
\bibinfo{author}{Bawuah, P.}, \bibinfo{author}{Markl, D.},
  \bibinfo{author}{Farrell, D.}, \bibinfo{author}{Evans, M.},
  \bibinfo{author}{Portieri, A.}, \bibinfo{author}{Anderson, A.},
  \bibinfo{author}{Goodwin, D.}, \bibinfo{author}{Lucas, R.},
  \bibinfo{author}{Zeitler, J.A.}, \bibinfo{year}{2020}.
\newblock \bibinfo{title}{{Terahertz-Based Porosity Measurement of
  Pharmaceutical Tablets: a Tutorial}}.
\newblock \bibinfo{journal}{Journal of Infrared, Millimeter, and Terahertz
  Waves} \bibinfo{volume}{41}, \bibinfo{pages}{450--469}.
\newblock \DOIprefix\doi{10.1007/s10762-019-00659-0}.
\bibitem[{Bawuah and Zeitler(2021)}]{Bawuah2021AdvancesReviewb}
\bibinfo{author}{Bawuah, P.}, \bibinfo{author}{Zeitler, J.A.},
  \bibinfo{year}{2021}.
\newblock \bibinfo{title}{{Advances in terahertz time-domain spectroscopy of
  pharmaceutical solids: A review}}.
\newblock \bibinfo{journal}{TrAC - Trends in Analytical Chemistry}
  \bibinfo{volume}{139}, \bibinfo{pages}{116272}.
\newblock \DOIprefix\doi{10.1016/j.trac.2021.116272}.
\bibitem[{Berntsson et~al.(1999)Berntsson, Burger, Folestad, Danielsson, Kuhn
  and Fricke}]{Berntsson1999EffectiveSpectrometry}
\bibinfo{author}{Berntsson, O.}, \bibinfo{author}{Burger, T.},
  \bibinfo{author}{Folestad, S.}, \bibinfo{author}{Danielsson, L.G.},
  \bibinfo{author}{Kuhn, J.}, \bibinfo{author}{Fricke, J.},
  \bibinfo{year}{1999}.
\newblock \bibinfo{title}{{Effective sample size in diffuse reflectance near-IR
  spectrometry}}.
\newblock \bibinfo{journal}{Analytical Chemistry} \bibinfo{volume}{71}.
\newblock \DOIprefix\doi{10.1021/ac980652u}.
\bibitem[{Berntsson et~al.(2002)Berntsson, Danielsson, Lagerholm and
  Folestad}]{Berntsson2002QuantitativeSpectroscopy}
\bibinfo{author}{Berntsson, O.}, \bibinfo{author}{Danielsson, L.G.},
  \bibinfo{author}{Lagerholm, B.}, \bibinfo{author}{Folestad, S.},
  \bibinfo{year}{2002}.
\newblock \bibinfo{title}{{Quantitative in-line monitoring of powder blending
  by near infrared reflection spectroscopy}}.
\newblock \bibinfo{journal}{Powder Technology} \bibinfo{volume}{123}.
\newblock \DOIprefix\doi{10.1016/S0032-5910(01)00456-9}.
\bibitem[{Chen et~al.(2017)Chen, Peng, Jiang, Zhao, Zhao and
  Zhu}]{Chen2017IsomersSpectroscopy}
\bibinfo{author}{Chen, W.}, \bibinfo{author}{Peng, Y.}, \bibinfo{author}{Jiang,
  X.}, \bibinfo{author}{Zhao, J.}, \bibinfo{author}{Zhao, H.},
  \bibinfo{author}{Zhu, Y.}, \bibinfo{year}{2017}.
\newblock \bibinfo{title}{{Isomers Identification of 2-hydroxyglutarate acid
  disodium salt (2HG) by Terahertz Time-domain Spectroscopy}}.
\newblock \bibinfo{journal}{Scientific Reports} \bibinfo{volume}{7}.
\newblock \DOIprefix\doi{10.1038/s41598-017-11527-z}.
\bibitem[{Claybourn et~al.(2007)Claybourn, Yang, Gradinarsky, Johansson and
  Folestad}]{Claybourn2007TerahertzApplicationsb}
\bibinfo{author}{Claybourn, M.}, \bibinfo{author}{Yang, H.},
  \bibinfo{author}{Gradinarsky, L.}, \bibinfo{author}{Johansson, J.},
  \bibinfo{author}{Folestad, S.}, \bibinfo{year}{2007}.
\newblock \bibinfo{title}{{Terahertz Spectroscopy for Pharmaceutical
  Applications}}.
\newblock \bibinfo{journal}{Handbook of Vibrational Spectroscopy} ,
  \bibinfo{pages}{1--14}\DOIprefix\doi{10.1002/9780470027325.s8914}.
\bibitem[{Ewing and Kazarian(2018)}]{Ewing2018RecentFormulations}
\bibinfo{author}{Ewing, A.V.}, \bibinfo{author}{Kazarian, S.G.},
  \bibinfo{year}{2018}.
\newblock \bibinfo{title}{{Recent advances in the applications of vibrational
  spectroscopic imaging and mapping to pharmaceutical formulations}}.
\newblock \bibinfo{journal}{Spectrochimica Acta - Part A: Molecular and
  Biomolecular Spectroscopy} \bibinfo{volume}{197}.
\newblock \DOIprefix\doi{10.1016/j.saa.2017.12.055}.
\bibitem[{Hisazumi et~al.(2012)Hisazumi, Watanabe, Suzuki, Wakiyama and
  Terada}]{Hisazumi2012UsingTablets}
\bibinfo{author}{Hisazumi, J.}, \bibinfo{author}{Watanabe, T.},
  \bibinfo{author}{Suzuki, T.}, \bibinfo{author}{Wakiyama, N.},
  \bibinfo{author}{Terada, K.}, \bibinfo{year}{2012}.
\newblock \bibinfo{title}{{Using terahertz reflectance spectroscopy to quantify
  drug substance in tablets}}.
\newblock \bibinfo{journal}{Chemical and Pharmaceutical Bulletin}
  \bibinfo{volume}{60}, \bibinfo{pages}{1487--1493}.
\newblock \DOIprefix\doi{10.1248/cpb.c12-00524}.
\bibitem[{H{\"{u}}bers(2008)}]{Hubers2008TerahertzReceivers}
\bibinfo{author}{H{\"{u}}bers, H.W.}, \bibinfo{year}{2008}.
\newblock \bibinfo{title}{{Terahertz heterodyne receivers}}.
\newblock \bibinfo{journal}{IEEE Journal on Selected Topics in Quantum
  Electronics} \bibinfo{volume}{14}, \bibinfo{pages}{378--391}.
\newblock \DOIprefix\doi{10.1109/JSTQE.2007.913964}.
\bibitem[{H{\"{u}}bers et~al.(2011)H{\"{u}}bers, Kimmitt, Hiromoto and
  Br{\"{u}}ndermann}]{Hubers2011TerahertzConsiderations}
\bibinfo{author}{H{\"{u}}bers, H.W.}, \bibinfo{author}{Kimmitt, M.F.},
  \bibinfo{author}{Hiromoto, N.}, \bibinfo{author}{Br{\"{u}}ndermann, E.},
  \bibinfo{year}{2011}.
\newblock \bibinfo{title}{{Terahertz spectroscopy: System and sensitivity
  considerations}}.
\newblock \bibinfo{journal}{IEEE Transactions on Terahertz Science and
  Technology} \bibinfo{volume}{1}, \bibinfo{pages}{321--331}.
\newblock \DOIprefix\doi{10.1109/TTHZ.2011.2159877}.
\bibitem[{Ierapetritou et~al.(2016)Ierapetritou, Muzzio and
  Reklaitis}]{Ierapetritou2016PerspectivesProcesses}
\bibinfo{author}{Ierapetritou, M.}, \bibinfo{author}{Muzzio, F.},
  \bibinfo{author}{Reklaitis, G.}, \bibinfo{year}{2016}.
\newblock \bibinfo{title}{{Perspectives on the continuous manufacturing of
  powder-based pharmaceutical processes}}.
\newblock \bibinfo{journal}{AIChE Journal} \bibinfo{volume}{62}.
\newblock \DOIprefix\doi{10.1002/aic.15210}.
\bibitem[{Jedvert et~al.(1998)Jedvert, Josefson and
  Langkilde}]{Jedvert1998QuantificationSpectroscopy}
\bibinfo{author}{Jedvert, I.}, \bibinfo{author}{Josefson, M.},
  \bibinfo{author}{Langkilde, F.}, \bibinfo{year}{1998}.
\newblock \bibinfo{title}{{Quantification of an active substance in a tablet by
  NIR and Raman spectroscopy}}.
\newblock \bibinfo{journal}{Journal of Near Infrared Spectroscopy}
  \bibinfo{volume}{6}.
\newblock \DOIprefix\doi{10.1255/jnirs.148}.
\bibitem[{Laske et~al.(2017)Laske, Paudel, Scheibelhofer, Sacher, Hoermann,
  Khinast, Kelly, Rantannen, Korhonen, Stauffer, De~Leersnyder, De~Beer,
  Mantanus, Chavez, Thoorens, Ghiotti, Schubert, Tajarobi, Haeffler, Lakio,
  Fransson, Sparen, Abrahmsen-Alami, Folestad, Funke, Backx, Kavsek, Kjell,
  Michaelis, Page, Palmer, Schaepman, Sekulic, Hammond, Braun and
  Colegrove}]{Laske2017AMolecules}
\bibinfo{author}{Laske, S.}, \bibinfo{author}{Paudel, A.},
  \bibinfo{author}{Scheibelhofer, O.}, \bibinfo{author}{Sacher, S.},
  \bibinfo{author}{Hoermann, T.}, \bibinfo{author}{Khinast, J.},
  \bibinfo{author}{Kelly, A.}, \bibinfo{author}{Rantannen, J.},
  \bibinfo{author}{Korhonen, O.}, \bibinfo{author}{Stauffer, F.},
  \bibinfo{author}{De~Leersnyder, F.}, \bibinfo{author}{De~Beer, T.},
  \bibinfo{author}{Mantanus, J.}, \bibinfo{author}{Chavez, P.F.},
  \bibinfo{author}{Thoorens, B.}, \bibinfo{author}{Ghiotti, P.},
  \bibinfo{author}{Schubert, M.}, \bibinfo{author}{Tajarobi, P.},
  \bibinfo{author}{Haeffler, G.}, \bibinfo{author}{Lakio, S.},
  \bibinfo{author}{Fransson, M.}, \bibinfo{author}{Sparen, A.},
  \bibinfo{author}{Abrahmsen-Alami, S.}, \bibinfo{author}{Folestad, S.},
  \bibinfo{author}{Funke, A.}, \bibinfo{author}{Backx, I.},
  \bibinfo{author}{Kavsek, B.}, \bibinfo{author}{Kjell, F.},
  \bibinfo{author}{Michaelis, M.}, \bibinfo{author}{Page, T.},
  \bibinfo{author}{Palmer, J.}, \bibinfo{author}{Schaepman, A.},
  \bibinfo{author}{Sekulic, S.}, \bibinfo{author}{Hammond, S.},
  \bibinfo{author}{Braun, B.}, \bibinfo{author}{Colegrove, B.},
  \bibinfo{year}{2017}.
\newblock \bibinfo{title}{{A Review of PAT Strategies in Secondary Solid Oral
  Dosage Manufacturing of Small Molecules}}.
\newblock \DOIprefix\doi{10.1016/j.xphs.2016.11.011}.
\bibitem[{Lyon et~al.(2002)Lyon, Lester, Lewis, Lee, Yu, Jefferson and
  Hussain}]{Lyon2002Near-infraredHomogeneity}
\bibinfo{author}{Lyon, R.C.}, \bibinfo{author}{Lester, D.S.},
  \bibinfo{author}{Lewis, E.N.}, \bibinfo{author}{Lee, E.},
  \bibinfo{author}{Yu, L.X.}, \bibinfo{author}{Jefferson, E.H.},
  \bibinfo{author}{Hussain, A.S.}, \bibinfo{year}{2002}.
\newblock \bibinfo{title}{{Near-infrared spectral imaging for quality assurance
  of pharmaceutical products: Analysis of tablets to assess powder blend
  homogeneity}}.
\newblock \bibinfo{journal}{AAPS PharmSciTech} \bibinfo{volume}{3}.
\newblock \DOIprefix\doi{10.1007/BF02830615}.
\bibitem[{Moradikouchi et~al.(2022)Moradikouchi, Spar{\'{e}}n, Folestad, Stake
  and Rodilla}]{Moradikouchi2022TerahertzTablets}
\bibinfo{author}{Moradikouchi, A.}, \bibinfo{author}{Spar{\'{e}}n, A.},
  \bibinfo{author}{Folestad, S.}, \bibinfo{author}{Stake, J.},
  \bibinfo{author}{Rodilla, H.}, \bibinfo{year}{2022}.
\newblock \bibinfo{title}{{Terahertz frequency domain sensing for fast porosity
  measurement of pharmaceutical tablets}}.
\newblock \bibinfo{journal}{International Journal of Pharmaceutics}
  \bibinfo{volume}{618}.
\newblock \DOIprefix\doi{10.1016/j.ijpharm.2022.121579}.
\bibitem[{Peng et~al.(2010)Peng, Peng, Jiang, Wei, Li and
  Tan}]{Peng2010AsymmetricCorrection}
\bibinfo{author}{Peng, J.}, \bibinfo{author}{Peng, S.}, \bibinfo{author}{Jiang,
  A.}, \bibinfo{author}{Wei, J.}, \bibinfo{author}{Li, C.},
  \bibinfo{author}{Tan, J.}, \bibinfo{year}{2010}.
\newblock \bibinfo{title}{{Asymmetric least squares for multiple spectra
  baseline correction}}.
\newblock \bibinfo{journal}{Analytica Chimica Acta} \bibinfo{volume}{683}.
\newblock \DOIprefix\doi{10.1016/j.aca.2010.08.033}.
\bibitem[{Peng et~al.(2018)Peng, Shi, Xu, Kou, Wu, Song, Ma, Guo, Liu and
  Zhu}]{Peng2018QualitativeSpectroscopy}
\bibinfo{author}{Peng, Y.}, \bibinfo{author}{Shi, C.}, \bibinfo{author}{Xu,
  M.}, \bibinfo{author}{Kou, T.}, \bibinfo{author}{Wu, X.},
  \bibinfo{author}{Song, B.}, \bibinfo{author}{Ma, H.}, \bibinfo{author}{Guo,
  S.}, \bibinfo{author}{Liu, L.}, \bibinfo{author}{Zhu, Y.},
  \bibinfo{year}{2018}.
\newblock \bibinfo{title}{{Qualitative and Quantitative Identification of
  Components in Mixture by Terahertz Spectroscopy}}.
\newblock \bibinfo{journal}{IEEE Transactions on Terahertz Science and
  Technology} \bibinfo{volume}{8}, \bibinfo{pages}{696--701}.
\newblock \DOIprefix\doi{10.1109/TTHZ.2018.2867816}.
\bibitem[{Plumb(2005)}]{Plumb2005ContinuousSet}
\bibinfo{author}{Plumb, K.}, \bibinfo{year}{2005}.
\newblock \bibinfo{title}{{Continuous processing in the pharmaceutical
  industry: Changing the mind set}}.
\newblock \bibinfo{journal}{Chemical Engineering Research and Design}
  \bibinfo{volume}{83}.
\newblock \DOIprefix\doi{10.1205/cherd.04359}.
\bibitem[{Pozar(2012)}]{Pozar2012MicrowaveEdition}
\bibinfo{author}{Pozar, D.M.}, \bibinfo{year}{2012}.
\newblock \bibinfo{title}{{Microwave Engineering, 4th Edition}}, in:
  \bibinfo{booktitle}{John Wiley {\&}Sons, Inc}, pp. \bibinfo{pages}{178--188}.
\bibitem[{Salim et~al.(2020)Salim, Fraser-Miller, Rziņ{\v{s}}, Sutton,
  Ramirez, Clulow, Hawley, Beilles, Gordon and
  Boyd}]{Salim2020Low-frequencyDigestion}
\bibinfo{author}{Salim, M.}, \bibinfo{author}{Fraser-Miller, S.J.},
  \bibinfo{author}{Rziņ{\v{s}}, K.B.}, \bibinfo{author}{Sutton, J.J.},
  \bibinfo{author}{Ramirez, G.}, \bibinfo{author}{Clulow, A.J.},
  \bibinfo{author}{Hawley, A.}, \bibinfo{author}{Beilles, S.},
  \bibinfo{author}{Gordon, K.C.}, \bibinfo{author}{Boyd, B.J.},
  \bibinfo{year}{2020}.
\newblock \bibinfo{title}{{Low-frequency raman scattering spectroscopy as an
  accessible approach to understand drug solubilization in milk-based
  formulations during digestion}}.
\newblock \DOIprefix\doi{10.1021/acs.molpharmaceut.9b01149}.
\bibitem[{Savitzky and Golay(1964)}]{Savitzky1964SmoothingProcedures}
\bibinfo{author}{Savitzky, A.}, \bibinfo{author}{Golay, M.J.},
  \bibinfo{year}{1964}.
\newblock \bibinfo{title}{{Smoothing and Differentiation of Data by Simplified
  Least Squares Procedures}}.
\newblock \bibinfo{journal}{Analytical Chemistry} \bibinfo{volume}{36}.
\newblock \DOIprefix\doi{10.1021/ac60214a047}.
\bibitem[{Shen et~al.(2008)Shen, Taday and
  Pepper}]{Shen2008EliminationSpectroscopy}
\bibinfo{author}{Shen, Y.C.}, \bibinfo{author}{Taday, P.F.},
  \bibinfo{author}{Pepper, M.}, \bibinfo{year}{2008}.
\newblock \bibinfo{title}{{Elimination of scattering effects in spectral
  measurement of granulated materials using terahertz pulsed spectroscopy}}.
\newblock \bibinfo{journal}{Applied Physics Letters} \bibinfo{volume}{92},
  \bibinfo{pages}{1--4}.
\newblock \DOIprefix\doi{10.1063/1.2840719}.
\bibitem[{Siegel(2004)}]{Siegel2004TerahertzMedicine}
\bibinfo{author}{Siegel, P.H.}, \bibinfo{year}{2004}.
\newblock \bibinfo{title}{{Terahertz technology in biology and medicine}}, in:
  \bibinfo{booktitle}{IEEE MTT-S International Microwave Symposium Digest}.
\newblock \DOIprefix\doi{10.1109/mwsym.2004.1338880}.
\bibitem[{Spar{\'{e}}n et~al.(2015)Spar{\'{e}}n, Hartman, Fransson, Johansson
  and Svensson}]{Sparen2015MatrixSpectroscopy}
\bibinfo{author}{Spar{\'{e}}n, A.}, \bibinfo{author}{Hartman, M.},
  \bibinfo{author}{Fransson, M.}, \bibinfo{author}{Johansson, J.},
  \bibinfo{author}{Svensson, O.}, \bibinfo{year}{2015}.
\newblock \bibinfo{title}{{Matrix effects in quantitative assessment of
  pharmaceutical tablets using transmission raman and near-infrared (NIR)
  Spectroscopy}}.
\newblock \bibinfo{journal}{Applied Spectroscopy} \bibinfo{volume}{69},
  \bibinfo{pages}{580--589}.
\newblock \DOIprefix\doi{10.1366/14-07645}.
\bibitem[{Svensson et~al.(2000)Svensson, Josefson and
  Langkilde}]{Svensson2000TheChemometrics}
\bibinfo{author}{Svensson, O.}, \bibinfo{author}{Josefson, M.},
  \bibinfo{author}{Langkilde, F.W.}, \bibinfo{year}{2000}.
\newblock \bibinfo{title}{{The synthesis of metoprolol monitored using Raman
  spectroscopy and chemometrics}}.
\newblock \bibinfo{journal}{European Journal of Pharmaceutical Sciences}
  \bibinfo{volume}{11}.
\newblock \DOIprefix\doi{10.1016/S0928-0987(00)00094-4}.
\bibitem[{Taday et~al.(2004)Taday, Van Der~Weide, Wood, Chamberlain, Roskos,
  Phillips, Newnham, Towrie and Appelquist}]{Taday2004ApplicationsSciences}
\bibinfo{author}{Taday, P.F.}, \bibinfo{author}{Van Der~Weide, D.},
  \bibinfo{author}{Wood, K.}, \bibinfo{author}{Chamberlain, M.},
  \bibinfo{author}{Roskos, H.}, \bibinfo{author}{Phillips, C.},
  \bibinfo{author}{Newnham, D.}, \bibinfo{author}{Towrie, M.},
  \bibinfo{author}{Appelquist, I.}, \bibinfo{year}{2004}.
\newblock \bibinfo{title}{{Applications of terahertz spectroscopy to
  pharmaceutical sciences}}.
\newblock \bibinfo{journal}{Philosophical Transactions of the Royal Society A:
  Mathematical, Physical and Engineering Sciences} \bibinfo{volume}{362},
  \bibinfo{pages}{351--364}.
\newblock \DOIprefix\doi{10.1098/rsta.2003.1321}.
\bibitem[{Townshend et~al.(2012)Townshend, Nordon, Littlejohn, Andrews and
  Dallin}]{Townshend2012EffectScattering}
\bibinfo{author}{Townshend, N.}, \bibinfo{author}{Nordon, A.},
  \bibinfo{author}{Littlejohn, D.}, \bibinfo{author}{Andrews, J.},
  \bibinfo{author}{Dallin, P.}, \bibinfo{year}{2012}.
\newblock \bibinfo{title}{{Effect of particle properties of powders on the
  generation and transmission of Raman scattering}}.
\newblock \bibinfo{journal}{Analytical Chemistry} \bibinfo{volume}{84},
  \bibinfo{pages}{4665--4670}.
\newblock \DOIprefix\doi{10.1021/ac203446g}.
\bibitem[{Trygg and Wold(2002)}]{Trygg2002OrthogonalO-PLS}
\bibinfo{author}{Trygg, J.}, \bibinfo{author}{Wold, S.}, \bibinfo{year}{2002}.
\newblock \bibinfo{title}{{Orthogonal projections to latent structures
  (O-PLS)}}.
\newblock \bibinfo{journal}{Journal of Chemometrics} \bibinfo{volume}{16}.
\newblock \DOIprefix\doi{10.1002/cem.695}.
\bibitem[{Vergote et~al.(2004)Vergote, De~Beer, Vervaet, Remon, Baeyens,
  Diericx and Verpoort}]{Vergote2004In-lineSpectroscopy}
\bibinfo{author}{Vergote, G.J.}, \bibinfo{author}{De~Beer, T.R.},
  \bibinfo{author}{Vervaet, C.}, \bibinfo{author}{Remon, J.P.},
  \bibinfo{author}{Baeyens, W.R.}, \bibinfo{author}{Diericx, N.},
  \bibinfo{author}{Verpoort, F.}, \bibinfo{year}{2004}.
\newblock \bibinfo{title}{{In-line monitoring of a pharmaceutical blending
  process using FT-Raman spectroscopy}}.
\newblock \bibinfo{journal}{European Journal of Pharmaceutical Sciences}
  \bibinfo{volume}{21}.
\newblock \DOIprefix\doi{10.1016/j.ejps.2003.11.005}.
\bibitem[{Wu et~al.(2008)Wu, Heilweil, Hussain and
  Khan}]{Wu2008ProcessApplication}
\bibinfo{author}{Wu, H.}, \bibinfo{author}{Heilweil, E.J.},
  \bibinfo{author}{Hussain, A.S.}, \bibinfo{author}{Khan, M.A.},
  \bibinfo{year}{2008}.
\newblock \bibinfo{title}{{Process analytical technology (PAT): Quantification
  approaches in terahertz spectroscopy for pharmaceutical application}}.
\newblock \bibinfo{journal}{Journal of Pharmaceutical Sciences}
  \bibinfo{volume}{97}, \bibinfo{pages}{970--984}.
\newblock \DOIprefix\doi{10.1002/jps.21004}.
\bibitem[{Yang et~al.(2021)Yang, Wu, Shi, Wu, Chen, Wu, Yang, Wang, Zeng and
  Peng}]{Yang2021QualitativeMethod}
\bibinfo{author}{Yang, Q.}, \bibinfo{author}{Wu, L.}, \bibinfo{author}{Shi,
  C.}, \bibinfo{author}{Wu, X.}, \bibinfo{author}{Chen, X.},
  \bibinfo{author}{Wu, W.}, \bibinfo{author}{Yang, H.}, \bibinfo{author}{Wang,
  Z.}, \bibinfo{author}{Zeng, L.}, \bibinfo{author}{Peng, Y.},
  \bibinfo{year}{2021}.
\newblock \bibinfo{title}{{Qualitative and Quantitative Analysis of Caffeine in
  Medicines by Terahertz Spectroscopy Using Machine Learning Method}}.
\newblock \bibinfo{journal}{IEEE Access} \bibinfo{volume}{9},
  \bibinfo{pages}{140008--140021}.
\newblock \DOIprefix\doi{10.1109/ACCESS.2021.3116980}.
\bibitem[{Zeitler et~al.(2007)Zeitler, Kogermann, Rantanen, Rades, Taday,
  Pepper, Aaltonen and Strachan}]{Zeitler2007DrugSpectroscopy}
\bibinfo{author}{Zeitler, J.A.}, \bibinfo{author}{Kogermann, K.},
  \bibinfo{author}{Rantanen, J.}, \bibinfo{author}{Rades, T.},
  \bibinfo{author}{Taday, P.F.}, \bibinfo{author}{Pepper, M.},
  \bibinfo{author}{Aaltonen, J.}, \bibinfo{author}{Strachan, C.J.},
  \bibinfo{year}{2007}.
\newblock \bibinfo{title}{{Drug hydrate systems and dehydration processes
  studied by terahertz pulsed spectroscopy}}.
\newblock \bibinfo{journal}{International Journal of Pharmaceutics}
  \bibinfo{volume}{334}, \bibinfo{pages}{78--84}.
\newblock \DOIprefix\doi{10.1016/j.ijpharm.2006.10.027}.

\end{thebibliography}

\end{document}


\section*{Supplementary data}

Fig. S1 shows the magnitude of the transmission coefficients ($S_{21}$) of the tablets and empty sample holder, having a short-open-load-through (SOLT) calibration, with the calibration planes at the waveguide flanges before the horn antennas, see Fig. 1.
\begin{figure}[h]
    \centering
    \includegraphics[scale=1.2]{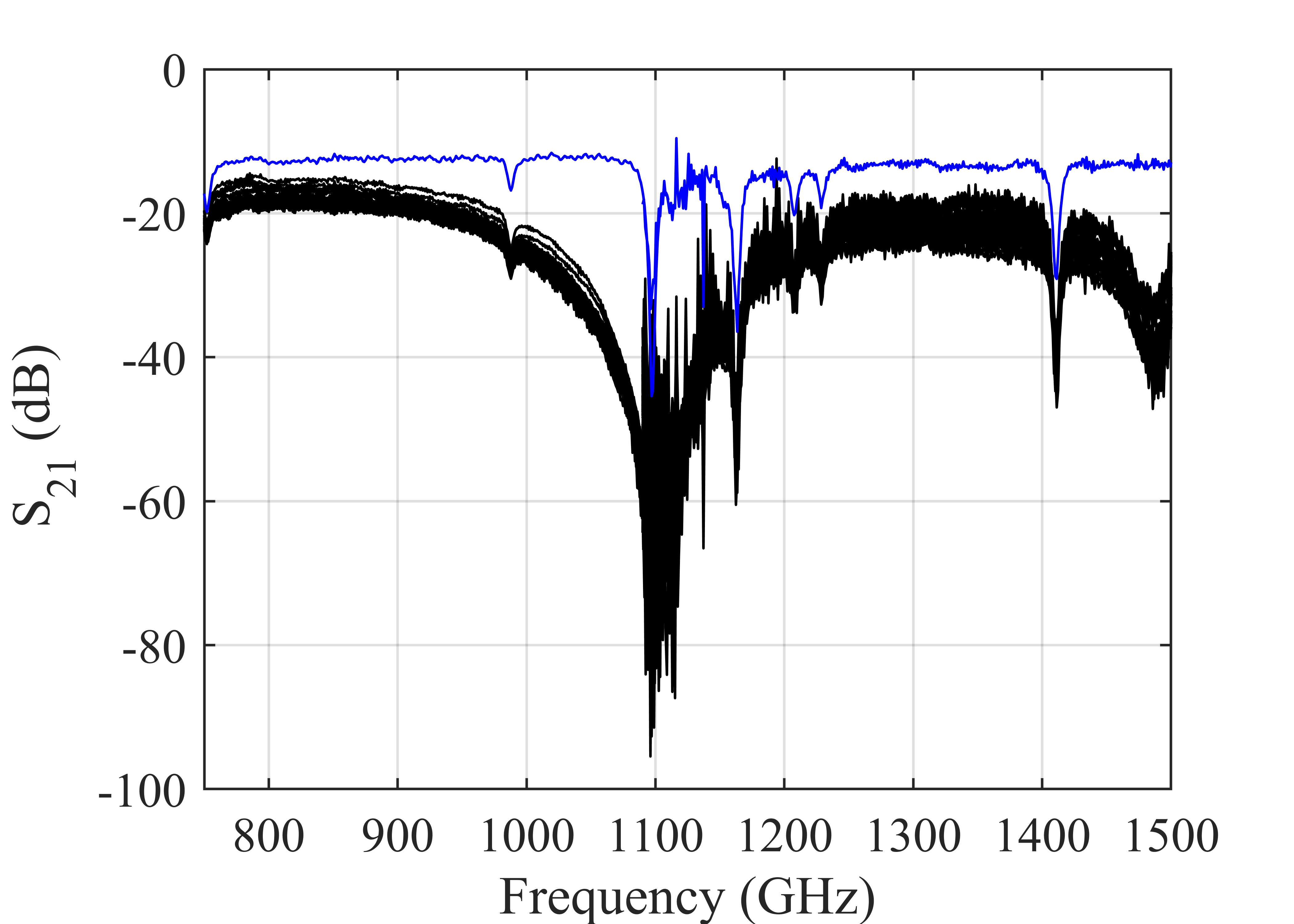}
\caption{$S_{21}$ parameters of the tablets listed in table 1, colored in black, and $S_{21}$ parameters of the empty sample holder, colored in blue. }
    \label{fig:sup 1}
\end{figure}